\documentstyle[11pt,newpasp,twoside, epsf]{article}
\def\edcomment#1{\iffalse\marginpar{\raggedright\sl#1\/}\else\relax\fi}
\reversemarginpar

\begin{document}
\title{Wide Field H-R diagrams of Local Group Dwarf Galaxies}
 \author{L. Rizzi and E.V. Held}
\affil{Osservatorio Astronomico di Padova, 
vicolo dell'Osservatorio 5, I-35122 Padova, Italy}
\author{I. Saviane}
\affil{European Southern Observatory, Casilla 19001, Santiago 19, Chile}
\author{Y. Momany}
\affil{Dipartimento di Astronomia, Universit\`a di Padova, 
vicolo dell'Osservatorio 1, I-35122 Padova, Italy}
\author{G. Clementini}
\affil{Osservatorio Astronomico di Bologna, Via Ranzani 1, I-40127 
Bologna, Italy}
\author{G. Bertelli}
\affil{Consiglio Nazionale delle Ricerche, CNR-GNA, Roma, Italy}
\begin{abstract}
We present new observations of Sextans and Leo~I dwarf galaxies, that
provide new insight on their stellar populations.
The large sample of stars measured in Sextans allowed us to obtain a
reliable estimate of the metal content of this galaxy, to study the
population spatial gradients, and to investigate the nature of the
large number of blue stragglers. In Leo~I, we discovered a significant
population of RR Lyrae variables.  The analysis of the pulsational
properties
of the detected variables suggests
that this galaxy is an Oosterhoff intermediate object, similar to the
LMC and to other dwarf galaxies in the Local Group.
\end{abstract}

\section{Introduction}
The work presented here is part of an ongoing project to determine the
star formation and chemical enrichment history of Local Group dwarf
galaxies. By means of CCD observations obtained with the WFI camera at
the MPG/ESO 2.2m telescope at La Silla, Chile, the different stellar
populations in the dwarf spheroidal (dSph) galaxies Sextans, Sculptor,
Leo I, Carina, Fornax and in the dwarf irregular NGC\,6822 have been
studied.  Here we highlight the main results obtained for the Sextans
and Leo I dwarf spheroidals.
\section{Stellar Populations in the Sextans dwarf spheroidal}
Discovered by Irwin et al. (\cite{irwinetal1990}), the Sextans dSph
galaxy is a perfect target for wide field observation, because of
its low stellar density and large extent (see Irwin \& Hatzidimitriou
\cite{irwin_hatzidimitriou}). 
Colour-magnitude diagrams of the galaxy were presented by Mateo et
al. (\cite{mateo1991}) and by Mateo, Fischer, \& Kreminski
(\cite{mateo1995}), the latter in a search for RR Lyrae
variables. Their diagrams showed, in addition to a predominantly old
population, a large number of stars brighter than the main sequence
turn off of a $\sim 10$~Gyr old population, 
temptatively identified as a
minor intermediate age component.  Both studies resulted in a
metallicity determination around [Fe/H]$=-1.6$.
Photometry in the Washington system was obtained by Geisler \&
Sarajedini (\cite{geisler1996}), yielding a metallicity of
[Fe/H]$=-2.0$.  A wide field colour-magnitude diagram was presented by
Pancino, Bellazzini, \& Ferraro (this conference), suggesting the
possible presence of a double population of stars at metallicities of
[Fe/H]$\sim-2.5$ and [Fe/H]$\sim-1.8$.
Spectroscopy of red giants was obtained by Suntzeff et
al. (\cite{suntzeff1993}) and by Shetrone, C\^ot\'e, \& Sargent
(\cite{shetrone2000}). Using the Ca~II infrared triplet, both studies
resulted in a value of about [Fe/H]$=-2.0$, with an intrinsic
abundance spread of 0.2 dex.

\begin{figure}
\plottwo{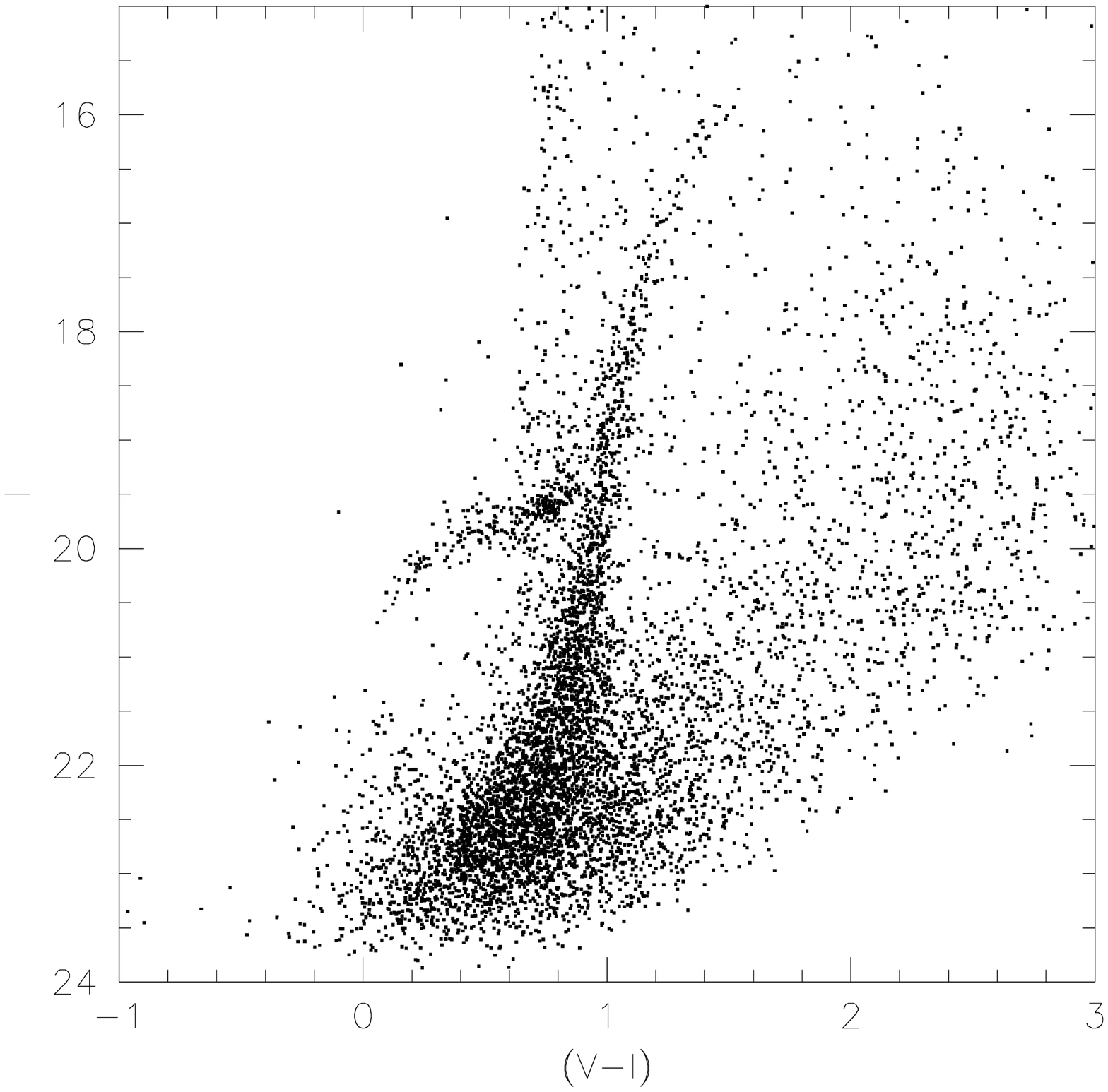}{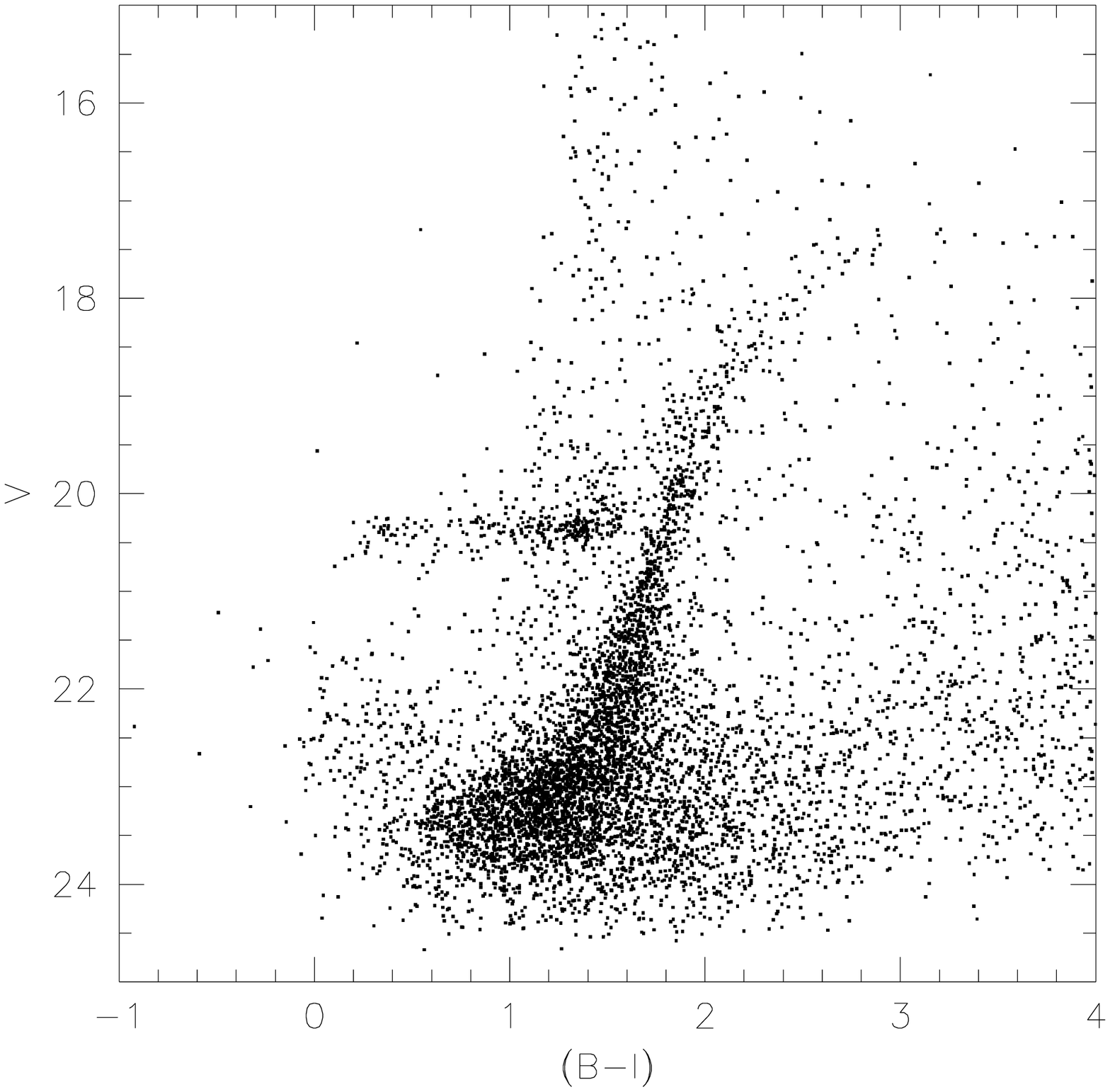}
\caption{
The $(V-I)$, $I$ and $(B-I)$, $V$ colour magnitude diagrams of the
Sextans dwarf spheroidal}
\label{fig1}
\end{figure}

Two questions were left open by previous studies: the first
is the determination of the metal content of this galaxy; the second
is the nature of the stars brighter than the main sequence turn
off. The observations presented here (Fig.\ref{fig1}) cover the
widest area ever observed ($\sim 0.5$ square degrees) and reach down
to magnitude V=24 (50\% completeness level).

\subsection{The metal content of Sextans}
These data allowed us to measure Sextans metallicity from the
comparison of the observed RGB with fiducial ridge lines of Galactic
globular clusters.
Using the method suggested by Saviane et al. (\cite{Saviane2000}),
based on iterative and simultaneous fitting of distance, metallicity,
and reddening, a value of ${\rm [Fe/H]}=-1.91\pm0.16$ was derived. The
intrinsic width of the RGB implies a metallicity dispersion of 0.2
dex. Both values are in agreement with previous spectroscopic
determination obtained with the Ca~II infrared triplet.

\subsection{The nature of the ``blue-stragglers''}
Mateo et al. (\cite{mateo1995}) suggested that the large number of
stars brighter than the main sequence turn off may be explained by the
presence of a minor intermediate age population.  Alternatively, these
could be old stars like those found in Galactic globular clusters, in
which case we expect them to be centrally concentrated.
To discriminate between these two hypotheses, we analyzed the spatial
distribution of these stars. No evidence of central concentration
could be found. Further, the intermediate age case was tested by means
of simulated diagrams obtained using the ZVAR code (G.~Bertelli 1997,
unpublished).

Our conclusion is that these stars most likely represent a secondary
star formation episode, since (a) the number of stars brighter than
the main sequence turn off is larger than in most globular clusters,
and (b) there are no signs of central concentration of these stars.
Indeed, Sextans is a very low concentration system that is unlikely to
form a large number of blue stragglers.  According to our analysis,
Sextans started forming stars at an early epoch ($\sim 13\div15$ Gyr
ago) and formed most of its stars in this first episode. A second
episode of star formation happened about 5 Gyr ago, in which up to 5
\% of the mass in stars was formed.
\section{Tracing the old population in Leo~I: the RR Lyrae variables}

With its predominantly young and intermediate age population and
seeming lack of an old component, Leo~I was long thought to be an
exception in the general picture of the Local Group dwarfs.

\begin{figure}[ht] 
\epsfxsize 7.5 cm 
\centerline{\epsfbox{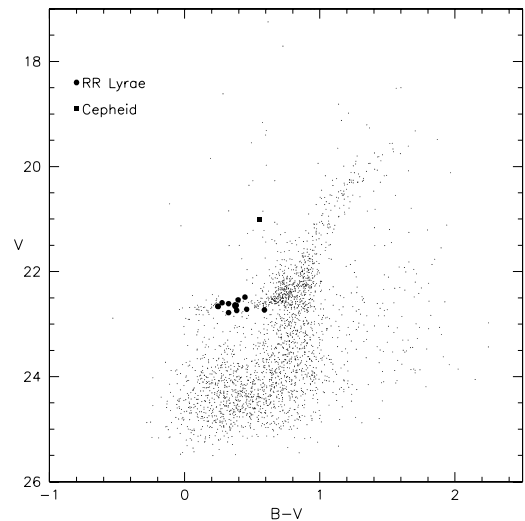}} 
\caption{CM diagram of Leo~I from CCD No. 6 
covering the outer region of the galaxy.
Filled circles represent the variable stars which has been plotted according to their average magnitudes and colors.
The brighter point is an anomalous Cepheid.}
\label{fig3}
\end{figure}

However, an old population has been revealed by a horizontal branch
extended from blue to red (Held et al. \cite{held2000}). Given the
presence of this extended HB and the low metallicity of the system
([Fe/H]$\sim -2$, Lee et al. \cite{lee1993}), one would naturally
expect to find RR Lyrae.
Indeed, a significant population of RR Lyrae variables was discovered
using a time series of 62 WFI camera images of the galaxy (Held et
al. \cite{clementini2001}).
Full coverage of the light variations and pulsation periods have been
obtained for 54 of the 74 candidates we found. 47 of them are Bailey
{\it ab}-type RR Lyrae and 7 are {\it c}-type.
Figure~\ref{fig3} shows the location of the detected variable stars in
the colour-magnitude diagram of Leo~I from the data of CCD No. 6.
The average period of the Leo~I RRab variables is
$\langle P_{ab}\rangle=0\fd60$, and the minimum period is $0\fd54$.
The pulsational properties of the RR Lyrae's qualify Leo~I as a system
intermediate between Oosterhoff I and Oosterhoff II clusters, similar
in this respect to other dwarf galaxies in the Local Group and to the
LMC.  Using Sandage (1993) formula, the average period of the {\it
ab}-type RR Lyrae's provides an independent estimate of the
metallicity for the old population: [Fe/H]$=-1.82$, although a
significant spread in metallicity seems to be present.
From the mean magnitude of the RR Lyrae's a distance modulus of
$(m-M)_0=22.04\pm0.12$ has been estimated using the metallicity value
derived in this work.  These results are in excellent agreement with
our recent metallicity and distance determinations on a different data
set (Momany et al. 2001, in preparation).

\end{document}